\documentclass[useAMS,usenatbib]{mn2e}
\input psfig.sty
\def\tem#1{\par\noindent
\hangindent6.5 mm\hangafter=0
\llap{#1\enspace}\ignorespaces}

\title[Masses of the components of HDE 226868/Cyg X-1]{Evolutionary constraints on the
masses of the components of HDE 226868/Cyg X-1 binary system}
\author[J. Zi\'o{\l}kowski]{J. Zi\'o{\l}kowski \thanks{E-mail:
jz@camk.edu.pl}\\N. Copernicus Astronomical Center, ul. Bartycka
18, 00-716 Warsaw, Poland}
\begin{document}

\date{Accepted 0000 December 00. Received 0000 December 00; in original form 0000 December 00}

\pagerange{\pageref{firstpage}--\pageref{lastpage}} \pubyear{2002}

\maketitle

\label{firstpage}

\begin{abstract}
Calculations carried out to model the evolution of HDE 226868,
under different assumptions about the stellar wind mass loss rate,
provide robust limits on the present mass of the star. It has to
be in the range 40 $\pm$ 5 M$_\odot$ if the distance to the system
is in the range 1.95 to 2.35 kpc and the effective temperature of
HDE 226868 in the range 30000 to 31000 K. Extending the possible
intervals of these parameters to 1.8 to 2.35 kpc and 28000 to
32000 K, one gets for the mass of the star the range 40 $\pm$ 10
M$_\odot$. Including into the analysis observational properties
such as the profiles of the emission lines, rotational broadening
of the absorption lines and the ellipsoidal light variations, one
can estimate also the mass of the compact component. It has to be
in the ranges 20 $\pm$ 5 M$_\odot$ and 13.5 $\div$ 29 M$_\odot$
for the cases described above. The same analysis (using the
evolutionary models and the observational properties listed above)
yields lower limit to the distance to the system of $\sim$ 2.0
kpc, if the effective temperature of HDE 226868 is higher than
30000 K. This limit to the distance does not depend on any
photometric or astrometric considerations.

\end{abstract}


\section{Introduction}

Cyg  X-1 was the first binary system in which the presence of a
black hole was suggested (Bolton 1972; Webster \& Murdin 1972).
For about a decade, it was the only object of that type. Many,
sometimes exotic, models and scenarios were devised to avoid the
presence of a black hole and to replace it with a neutron star.
With the advent of subsequent black hole candidates (LMC X-3,
A620$-$00), the motivation for such efforts substantially
diminished. At present, there are no doubts about the presence of
a black hole in the system. However, in spite of three decades of
the investigations, there is still substantial uncertainty
concerning the masses of both components. The mass function is
known rather precisely. Its most recent value was given by Gies et
al. (2003): $f(M_x) = 0.251 \pm 0.007$ M$_\odot$. There are also
no doubts that the optical component must be close to filling its
Roche lobe. Gies \& Bolton (1986a, 1986b), analyzing the emission
lines of the stellar wind, the rotational broadening of the
absorption lines of the optical component and the photometric
V-band light curve, found that the fill-out factor must be greater
than 0.9 and its best value is in the range 0.95 to 1.The value of
the inclination of the orbit is less certain. From the analysis
mentioned above (rotational broadening of the absorption lines of
the optical component and the ellipsoidal light variations in the
V-band light curve), the authors estimated the inclination of the
orbit to be $i = 33 \pm 5^{\rm o}$. On the other hand, from
polarimetric measurements (in three colours) Dolan \& Tapia (1989)
found the inclination $i = 62_{-37}^{\hspace*{0.8ex}+5}$ $^{\rm
o}$, which, while different, is not inconsistent with the value of
Gies \& Bolton.

Paczy\'nski (1974) has shown that basing only on the lack of the
X-ray eclipses and the fact that the optical component cannot be
larger than its Roche lobe, one can obtain lower limits to the
masses of both components as functions of the distance to the
system. Paczy\'nski found that the mass of the compact component
has to be larger than 3.2 M$_\odot$ if the distance is greater
than 1.3 kpc. He noticed also that, basing on stellar evolution
consideration, the distance cannot be larger than about 2.1 kpc.
The distance to Cyg X-1 is not very well established, but,
certainly, it is greater than 1.3 kpc. Margon, Bowyer \& Stone
(1973) and Bregman et al. (1973) analyzed interstellar reddening
in the direction of HDE 226868 and found that its distance is
about 2.5 kpc. Wu et al. (1982), from a study of the UV colors of
HDE 226868, concluded that its distance must be greater than 1.9
kpc. Bregman's value was confirmed by Ninkov, Walker \& Young
(1987), who got the distance of 2.5 $\pm$ 0.3 kpc from the
equivalent width of the H$_\gamma$ line. HDE 226868 lies only
$1^{\rm o}$ from the center of NGC 6871 (the core of the Cyg OB3
association) and seems to be its member. However, the estimates of
the distance to the association (or its core) gave mixed results.
Crawford, Barnes \& Warren (1974) used three methods (UBV
photometry, the H-R diagram fitting and the equivalent width of
the H$_\beta$ line) and obtained the distance to the association
of about 2 kpc. Humphreys (1978) estimated the photometric
distance to Cyg OB3 to be about 2.3 kpc. Janes \& Adler (1982)
quote in their catalogue the distance to NGC 6871 as only 1.8 kpc.
The value of 1.8 kpc as the distance to Cyg OB3 was obtained again
by Garmany \& Stencel (1992), who used the H-R diagram fitting
method. The discrepancy between the estimates of the distances to
HDE 226868 and to Cyg OB3 cast the doubt on the membership of Cyg
X-1 in NGC 6871. Fortunately, the extensive photometry and
spectroscopy by Massey, Johnson \& DeGioia-Eastwood (1995) seemed
to clarify the situation. They estimated the distance to NGC 6871
as 2.14 $\pm$ 0.07 kpc, which was in reasonable agreement with the
independent (H$_\gamma$ width) estimate of the distance to HDE
226868 (Ninkov et al. 1987). Let us note that a very similar value
of the distance ($d \approx$ 2.15 kpc) was derived also by Gies \&
Bolton (1986a) as a byproduct of their careful analysis of a large
set of observational data for HDE 226868/Cyg X-1 binary system.
Their result (see the point $M_{\rm x} = 16$ M$_\odot$ and $M_{\rm
opt} = 33$ M$_\odot$ in their fig. 10) was obtained quite
independently of any photometric considerations. We should also
note, however, that while Herrero et al. (1995) do not discuss
explicitly the value of the distance in their atmosphere modelling
paper, their final model implies the value $\sim$ 1.8 kpc.
Finally, the question of Cyg X-1 membership in NGC 6871 was
definitely solved in recent paper by Mirabel \& Rodriguez (2003).
Comparing the high precision  VLBI astrometry for Cyg X-1 and
Hipparcos astrometry for the members of Cyg OB3, the authors
convincingly demonstrated that Cyg X-1 shares, quite precisely,
common proper motion with Cyg OB3 association. The velocity of Cyg
X-1 with respect to the Sun is 70 $\pm$ 3 km/s. The relative space
velocity of Cyg X-1 with respect to Cyg OB3 is only 9 $\pm$ 2
km/s, which is typical of the random velocities of stars in
expanding associations (Blaauw 1991).

After the work by Massey et al. (1995), several more estimates of
distances to both Cyg OB3 and HDE 226868 were made and the results
still exhibit a substantial scatter. Malysheva (1997), using
different photometry than Massey et al. (1995), obtained for the
distance to Cyg OB3 again $d \approx$ 1.8 kpc. Dambis, Mel'nik \&
Rastorguev (2001) calculated the trigonometric distance to Cyg OB3
as a median value of the Hipparcos distances to the 18 individual
member stars and found $d$ = 2.3 (+1.4,$-$0.6) kpc. In the case of
Cyg OB3 their trigonometric distance agreed very well with the
photometric distance of Blaha \& Humphreys (1989), although they
found that, typically, Hipparcos trigonometric distances to OB
associations are by $\sim$ 12\% smaller than photometric distances
of Blaha \& Humphreys. A recent photometric distance estimate made
by Miko{\l}ajewska (2002, private communication) gave the result
$d$ = 1.95 $\pm 0.10$ kpc. Lestrade et al. (1999) used the VLBI
astrometry to estimate the distance to Cyg X-1 and obtained $d$ =
1.0 $\div$ 2.3 kpc.

All this extensive effort leads to the conclusion that the
distance to HDE 226868/Cyg X-1 binary system cannot be
significantly different from 2 kpc. Therefore, I believe that the
best choice will be to follow Massey et al. (1995) and use $d$ =
2.15 $\pm$ 0.07 $(1 \sigma$ error) or $\pm$ 0.2 $(3 \sigma$ error)
kpc as the distance to Cyg X-1. As we shall see at the end of the
paper, this choice will be, quite independently, supported by
evolutionary considerations. The above value is used throughout
the rest of this paper (except when I assume the distance to be a
free parameter). In those parts of the discussion where the
distance is assumed to be a free parameter, I shall consider the
value 1.8 kpc as a reasonable lower limit to the distance of Cyg
X-1.

\section{The  Model Independent Lower Limits to the Components Masses}

In this section, I shall repeat Paczy\'nski's (1974) analysis,
using the present day data. This analysis is based only on solid
observational facts and is, therefore, model independent. Let us
remind that these solid facts include: (i) the value of the mass
function, (ii) the lack of the X-ray eclipses, (iii) the obvious
fact that the optical component cannot be larger than its Roche
lobe, and (iiii) the spectral type and the photometry of the
optical component (used to derive its effective temperature and
the bolometric correction).

The most recent determination of the mass function is $f(M_x) =
0.251 \pm 0.007$ M$_\odot$ (Gies et al. 2003), which differs only
slightly from the earlier values $f(M_x) = 0.244 \pm 0.005$
M$_\odot$ (Brocksopp et al. 1999) and $f(M_x) = 0.252$ M$_\odot$
(Gies \& Bolton 1982). The corresponding projected radius of the
optical component orbit is $a_1$sin$i = 8.36 \pm 0.08$ R$_\odot$.
The spectral type of HDE 226868 was determined by Walborn (1973)
as O9.7 Iab. This classification was confirmed by Gies \& Bolton
(1986a), by Ninkov et al. (1987) and by Herrero et al. (1995) and
was used in most of the literature. On the other hand, Massey et
al. (1995) obtained, from their new massive spectroscopy of OB
associations, the somewhat earlier spectral type - ON9 Ifa+.
According to the calibration of effective temperatures for late
O-type supergiants derived by Vacca, Garmany \& Shull (1996,
hereafter VGS), one has $T_{\rm e}$ = 32740 K for O9 Ia star and
extrapolation for O9.7 Ia gives $T_{\rm e}$ = 30690 K. On the
other hand, Herrero et al. (1995) used atmosphere models to
reproduce the observed spectrum of HDE 226868 and obtained $T_{\rm
e}$ = 32000 K as the best fit for the effective temperature of
this star.

Both VGS and Herrero et al. calibrations were based on pure H-He
models of the atmospheres that neglected the effects of metallic
lines blanketing. It is well known by now, that taking into
account the line blanketing leads to the lower effective
temperature for a given spectral type (Martins et al. 2002,
Repolust et al. 2004, Markova et al. 2004). The most recent
calibration of effective temperatures for O-type supergiants
(accounting for the line blanketing effects) was derived by
Markova et al. (2004). Their scale gives generally lower effective
temperatures than VGS scale (by up to 10 000 K for the earliest
spectral types). However, for the spectral type O9.7 I, both
temperature scales converge to the value $T_{\rm e}$ = 30700 K. It
seems that, at the present state of knowledge, the most reasonable
approach would be to assume that $T_{\rm e}$ = 30700 K is the best
estimate of the effective temperature of HDE 226868. To take into
account the uncertainty of this estimate, in further discussion I
shall consider a broad interval of $T_{\rm e} = 28000 \div 32000$
K, as the range of the possible values of the effective
temperature of HDE 226868.

As the effective temperature-bolometric correction relation is not
affected by the line blanketing (Martins et al. 2002), it is
possible to take this relation from VGS tables. For the effective
temperature 30700 K, the appropriate value of the bolometric
correction is $B.C.$ = $-$3.06 . The corresponding unreddened
colour index should be $(B-V)_{\rm o} = -0.26$ (Schmidt-Kaler
1982). The most recent photometry of HDE 226868 by Massey et al.
(1995) gives $V = 8.81$ and $B-V = 0.83$. Therefore the reddening
is $E_{\rm B-V} = 1.09$ and the unreddened $V$ magnitude is
$V_{\rm o} = 5.43$ (following Massey et al., I adopt $R_{\rm V} =
3.1$). With the effective temperature and the bolometric
correction given above, we can then express the radius and
luminosity of HDE 226868 as functions of the distance:
\begin{eqnarray}
R_{\rm opt}/ R_\odot = 10.59\hspace*{.5ex} (d/1 {\rm kpc}),
\end{eqnarray}
\begin{eqnarray}
L_{\rm opt}/ L_\odot = 8.95 \times 10^4\hspace*{.5ex} (d/1 {\rm
kpc})^2.
\end{eqnarray}

Following the procedure of Paczy\'nski (1974), we can now
calculate the lower limits to the masses of both components as
functions of the distance to the system (under the assumptions
listed at the beginning of this section). The results of these
calculations are given in the first part of Table 1.

\begin{table}
%\centerline{\bf Tab. 1 $-$ Parameters of the components of HDE
%226868/Cyg X-1} \centerline{\bf binary system as functions of the
%assumed distance d} \vspace{5mm}
%\nopagebreak
\centering

 \begin{minipage}{75mm}
  \caption{Parameters of the components of HDE 226868/Cyg X-1
   binary system as functions of the assumed distance $d$ for
   three values of the effective temperature of the optical component.}

\begin{tabular}{|r|r|r|r|r|}
\hline \multicolumn{1}{|c|}{$d$}&\multicolumn{1}{|c|}{$R_{\rm
opt}$}&\multicolumn{1}{|c|}{$M_{\rm
opt,min}$}&\multicolumn{1}{|c|}{ $M_{\rm x,min}$}&\multicolumn{1}{
|c|}{log($L$/L$_{\odot}$)}\\
\multicolumn{1}{|c|}{[kpc]}&\multicolumn{1}{|c|}{[R$_{\odot
}$]}&\multicolumn{1}{|c|}{[M$_{\odot}$]}&\multicolumn{1}{|c|}{[M$_{\odot
}$]}&\\ \hline\\ \multicolumn{5}{|c||}{$T_{\rm e} = 30700$ K}\\
\hline\\ 1.80&19.1&20.4&6.&5.46\\ 1.95&20.7&25.4&7.4&5.53\\
2.08&22.0&30.2&8.3&5.59\\ 2.15&22.8&33.0&8.7&5.62\\
2.22&23.5&36.0&9.2&5.65\\ 2.35&24.9&41.9&10.2&5.69\\
2.50&26.5&49.5&11.3&5.75\\ 2.70&28.6&60.9&12.9&5.82\\ \hline\\
\multicolumn{5}{|c||}{$T_{\rm e} = 28000$ K}\\ \hline\\
1.80&19.7&22.3&6.8&5.33\\ 1.95&21.3&27.6&7.8&5.40\\
2.08&22.8&32.9&8.7&5.46\\ 2.15&23.5&35.9&9.2&5.49\\
2.22&24.3&39.2&9.&5.51\\ 2.35&25.7&45.7&10.8&5.56\\
2.50&27.3&53.9&12.0&5.62\\ 2.70&29.5&66.3&13.7&5.68\\ \hline\\
\multicolumn{5}{|c||}{$T_{\rm e} = 32000$ K}\\ \hline\\
1.80&18.7&19.5&6.3&5.52\\ 1.95&20.3&24.1&7.2&5.59\\
2.08&21.6&28.7&8.0&5.64\\ 2.15&22.4&31.4&8.5&5.67\\
2.22&23.1&34.2&8.9&5.70\\ 2.35&24.4&39.9&9.9&5.75\\
2.50&26.0&47.1&11.0&5.80\\ 2.70&28.1&57.9&12.5&5.87\\ \hline\\
\end{tabular}
\end{minipage}
\end{table}

Let us notice that the changes, comparing with Paczy\'nski's Table
1, are mostly due to substantial change in the effective
temperature (Paczy\'nski took $T_{\rm e}$ = 25000 K)  and the
corresponding change in the bolometric correction for HDE 226868.
Let us also notice that the effects of the changes in the
effective temperature and the bolometric correction, to a large
degree, cancel each other when we consider the radius of HDE
226868 and the lower limits to the masses of both components. They
introduce, however, the dramatic (by a factor of $\sim 2.5$)
increase of the luminosity of HDE 226868 (for a given distance).
As a result, HDE 226868 appears to be a very bright star: its
optical luminosity is $\sim$ 1 $\div$ 2 $\times$ 10$^{39}$ erg/sec
(by two orders of magnitude higher than the typical X-ray
luminosity of Cyg X-1).

Undoubtedly, we know now the calibration of effective temperatures
for O-type supergiants much better than 30 years ago. However,
still, there remains some uncertainty, clearly demonstrated by the
recent major revision introduced by accounting for the line
blanketing effects (as discussed earlier). To investigate the
effect of this uncertainty, I assumed that the effective
temperature of HDE 226868 lies in the range 28000 $\div$ 32000 K
and calculated the corresponding lower limits to the masses of
both components (using the modified versions of eqs. (1)$-$(2)).
The results of these calculations are given in the second and the
third part of Table 1. All sets of results are illustrated in Fig.
1.

Comparing the both sections of the table (or looking at the Fig.
1), we may notice that the uncertainty of the effective
temperature does not affect significantly the lower limits to the
masses.

Summarizing the considerations of this section, we may state that
the masses of the compact and the optical components must be
greater than $\sim 8$ M$_\odot$ and $\sim 29$ M$_\odot$,
respectively (if the distance to the system is in the range 2.15
$\pm$ 0.07 kpc). Let us remind that these values are model
independent and therefore very hard to contest. If one wants to be
more conservative and uses the $3 \sigma$ error as the uncertainty
of the distance estimate, then these lower limits become $\sim 7$
M$_\odot$ and $\sim 24$ M$_\odot$, respectively. Even adopting the
smallest value ever claimed for the distance ($d$ = 1.8 kpc), one
still obtains the limits $\sim 6$ M$_\odot$ and $\sim 19$
M$_\odot$, respectively. One should add that our method gives only
approximate estimates, since (as noted by Gies 2004, private
communication) it assumes isotropic optical emission, which is not
the case for tidally distorted star. The resulting inaccuracies
should not be, however, significant.

The lower limits, presented in Table 1 and Fig. 1, become the
actual values if the fill-out factor is equal 1.0 and the
inclination of the orbit corresponds to the grazing eclipse
orientation. The fill-out factor cannot be much different from 1.0
(Gies \& Bolton 1986a,b). However, the inclination is the source
of larger uncertainty. If the inclination is close to $62^{\rm o}$
(as suggested by the polarimetry, Dolan \& Tapia 1989), then we
are close to the grazing eclipse situation and the masses are
close to the lower limits obtained in this section, i.e. $8 \div
9$ M$_\odot$ and $30 \div 36$ M$_\odot$, respectively. If (as it
seems more likely $-$ see the discussion in Section 5), the
inclination is rather close to $33^{\rm o}$ (advocated by Gies \&
Bolton 1986a and still not excluded by polarimetry), then the
masses are substantially higher and probably close to the values
suggested by Gies \& Bolton ($16 \pm 5$ M$_\odot$ and $33 \pm 9$
M$_\odot$, respectively). I shall return to this point in further
discussion.

\vspace{0.5cm}

\begin{figure}
%\epsfysize=10cm % fix the y-dimension and scales x-dim. to y-dim.
%\epsfxsize=10cm % fix the x-dimension and scales y-dim. to x-dim.
% Feel free to do the choice you prefer but do not exceed the x-dimension
% of the text lines
 %\includegraphics[width=84mm]{m-d.eps}
 %\includegraphics[width=84mm]{mx-d.eps}
%\hspace{0.2cm}\epsfbox{m-d.eps} %for centering: act on hspace argument
%\hspace{0.0cm}\epsfbox{mx-d.eps} %for centering: act on hspace argument
\centerline{\psfig{file=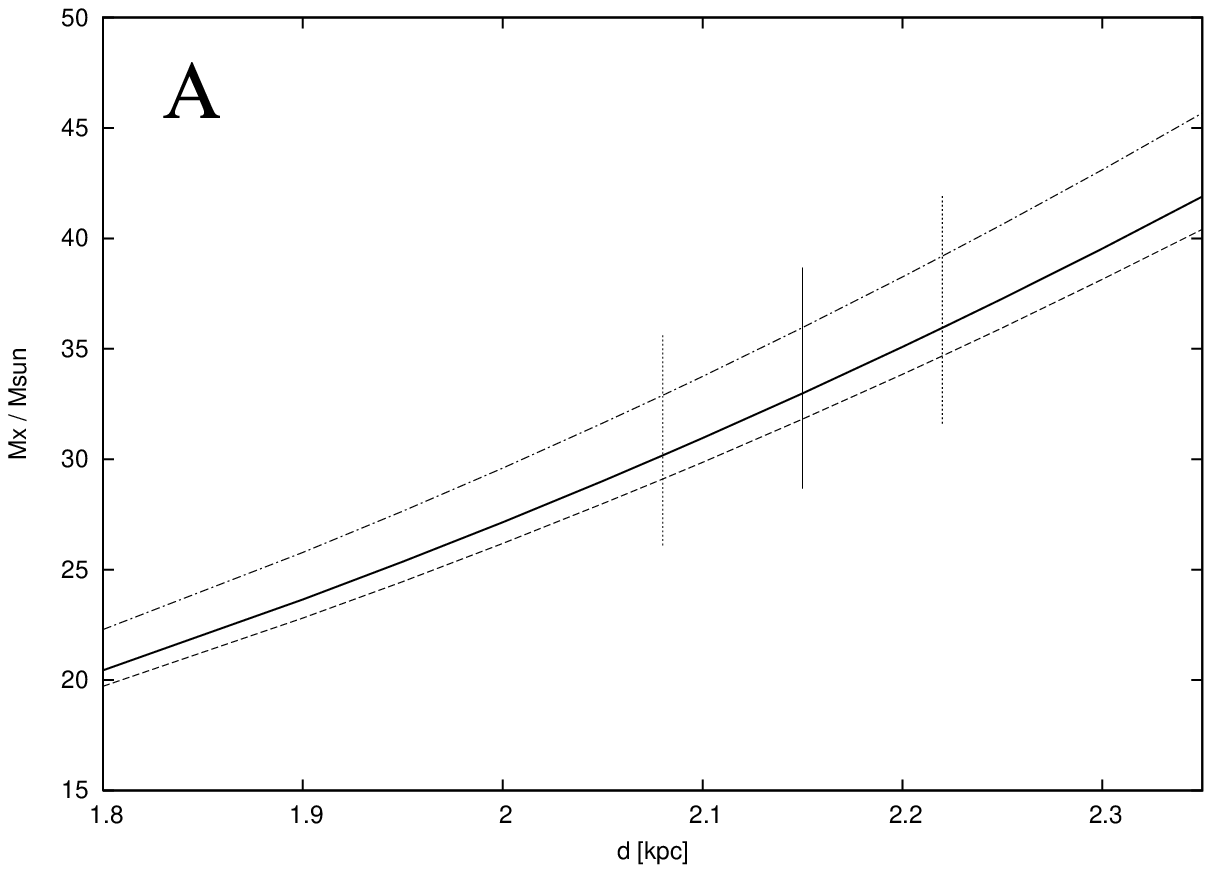,width=8.7cm}}
\centerline{\psfig{file=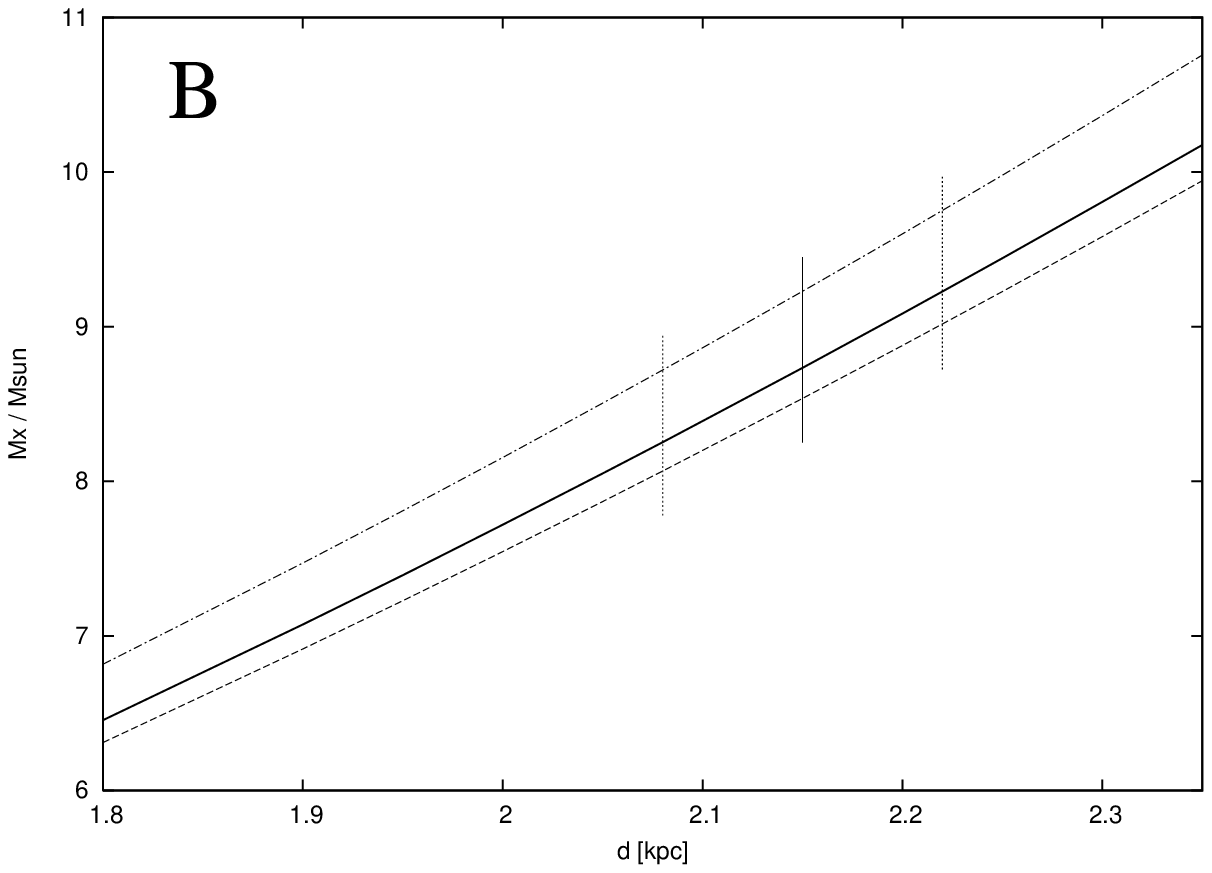,width=8.7cm}}

\caption[h]{{\footnotesize The lower limits to the masses of the
optical (A) and compact (B) components as functions of the assumed
distance, $d$. Thick solid lines correspond to the most likely
value of the effective temperature of HDE 226868 ($T_{\rm e}$ = 30
700 K). The broken lines and the dash-dotted lines correspond to
the effective temperatures 28 000 K and 32 000 K, respectively.
Thin solid vertical line indicates the most likely value of the
distance ($d$ = 2.15 kpc). The broken vertical lines indicate $\pm
1 \sigma$ errors in the distance estimate. The $\pm 3 \sigma$
error range corresponds to the distance interval 1.95 $\div$ 2.35
kpc.}}
  \label{f1}
\end{figure}

\section{The  Evolutionary Status of HDE 226868}

Quite independent constraints on the masses may be obtained from
analysis of the evolutionary status of the optical component.
First, let us note that we know relatively well its luminosity. At
the distance of 2.15 $\pm$ 0.07 kpc, the luminosity must be
$M_{\rm bol} \approx - 9.29 \pm 0.07$ (equivalent to
log($L$/L$_{\odot}$) = 5.62 $\pm$ 0.03 in Table 1). Including the
uncertainty of the effective temperature and assuming $3 \sigma$
error in the distance estimate, the range for the luminosity
becomes $M_{\rm bol} \approx - 8.8 \div - 9.6$ (equivalent to
log($L$/L$_{\odot}$) = 5.40 $\div$ 5.75 in Table 1). The real
range is somewhat wider because one should include also the errors
in $E_{\rm B-V}$ (about $\pm$ 0.05) and $R_{\rm V}$ (about $\pm$
0.1) estimates. The final range becomes therefore $M_{\rm bol}
\approx - 8.5 \div - 9.9$. One may note that this range
corresponds to $M_{\rm V} \approx - 5.7 \div - 6.7$, which agrees
quite well with $M_{\rm V} \approx - 6.5 \pm 0.2$ obtained by
Ninkov et al. (1987) from the equivalent width of the H$_\gamma$
line.

Second, let us consider the evolutionary phase of HDE 226868. Even
without the specific evolutionary calculations, it is relatively
straightforward to argue that it must be a core hydrogen burning
configuration. It  comes from the fact that HDE 226868 must be in
a relatively stable phase of its evolution. We observe no
measurable variations of the orbital period of the system. The $3
\sigma$ upper limit for the relative change of the orbital period
$\mid$d ln $P$/d $t\mid$ is $\sim 2 \div 3 \times 10^{-5}$
yr$^{-1}$ (Gies \& Bolton 1982). This implies that the present
evolutionary timescale of HDE 226868 must be substantially longer
than $\sim 3 \div 5 \times 10^4$ yr. This excludes the possibility
of the post-main sequence evolution, as the expected evolutionary
time scale is, in that case, of the order of $10^3$ yr.
Zi\'o{\l}kowski (1977), investigating the post-main sequence
evolution of a $25$ M$_\odot$ star (of similar radius but smaller
luminosity than HDE 226868) approaching its Roche lobe, found that
its radius was increasing from the fill-out factor equal 0.95 to
1.00 in only $\sim$ 600 yr. After filling the Roche lobe, it took
only $\sim$ 200 yr for the mass transfer rate to exceed the
Eddington limit by 3 orders of magnitude. For HDE 226868, which is
more luminous and, probably, more massive ($35 \div 45$ M$_\odot$,
as we shall see later), the relevant time scales should be even
shorter. If HDE 226868 were expanding that fast (5\% radius
increase in less than 600 yr), we should observe the noticeable
rise of its stellar wind strength and of the X-ray luminosity of
the system (both parameters are sensitive to the fill-out factor
if it is close to 1). Moreover, the chance of observing the system
during such a short time window (just few hundred years, before it
gets extinguished as an X-ray source due to hyper-Eddington
accretion) is very small. Therefore, we have to assume  that HDE
226868 (similarly as the optical components of other massive X-ray
binaries - see Zi\'o{\l}kowski 1977) must still burn hydrogen in
its core.

Exactly the same conclusion may be obtained directly as a unique
outcome of the numerical modelling of the evolution of HDE 226868.
This topic will be discussed in the next section.

\section{The  Evolutionary Calculations for HDE 226868}

\subsection{The  general description}

I computed evolutionary tracks for core hydrogen burning phase of
stars with the initial masses in the range $40 \div 80$ M$_\odot$.
The Warsaw evolutionary code developed by Bohdan Paczy\'nski and
Maciek Koz{\l}owski and kept updated by Ryszard Sienkiewicz was
used. The initial chemical composition $X$=0.7 and $Z$=0.03 is
adopted. The opacity tables incorporating OPAL opacities (Iglesias
\& Rogers 1996) as well as molecular and grain opacities
(Alexander \& Ferguson 1994) were used. The nuclear reaction rates
are those of Bahcall \& Pinsonneault (1995). The equation of state
used was Livermore Laboratory OPAL (Rogers, Svenson \& Iglesias
1996). I neglected the semiconvective mixing, as it is not
important during the evolutionary phase considered (most of the
models of interest had central hydrogen content $X_{\rm c} \ga
0.2$). Similarly any overshooting at the border of the convective
core was neglected (it is even less important).

The calculations were carried out under the assumption that the
evolution starts from the homogeneous configurations. It means
that the consequences of the fact that some of the matter of the
star, possibly dumped from the progenitor of the present black
hole, could have somewhat modified chemical composition, were
neglected. It means also, that the consequences of the fact that
some nuclear evolution (hydrogen burning) could, possibly, take
place while the mass of the star was smaller (prior to the mass
transfer) were neglected as well. It seems that both
simplifications do not alter significantly the outcome of the
evolutionary calculations. I shall return to this point in later
discussion.

\subsection{The  stellar wind mass loss}

The most uncertain element of the calculations of the early
evolution of massive stars is the mass loss due to stellar wind.
The uncertainty of the estimate of its rate is the single most
important factor influencing the outcome of the calculations. The
observations seem to indicate that there is a substantial scatter
of the mass loss rates (typically, by a factor of two, but
sometimes this factor can reach up to five) among the stars of
similar luminosities and effective temperatures (see e.g. de
Jager, Nieuwenhuijzen \& Van Der Hucht 1988). I decided to use the
prescription given by Nieuwenhuijzen \& de Jager (1990) in the
form of the formula derived by Hurley, Pols \& Tout (1999,
hereafter HPT). Bearing in mind that the formula gives the mass
loss rate estimate with the accuracy that is probably not better
than within a factor of two, I introduced the multiplying factor
$f_{\rm SW}$ applied to HPT formula. For each initial mass of the
star  three evolutionary sequences were calculated with the value
of the parameter $f_{\rm SW}$ equal, in sequence, 0.5, 1 and 2. In
this way, the uncertainty of the theory of evolution could be,
hopefully, taken into account. It might be interesting to note, at
this point, the astonishingly good agreement between the HPT
formula and the observational determination of the mass loss rate
for HDE 226868: for the most likely values of the parameters of
the star, the HPT value agrees with the observed one ($\dot{M} = -
2.6 \times 10^ {-6}$ M$_\odot$/yr, Gies et al. 2003, see below) to
an accuracy of about 5\% (this is, definitely, better than the
precision of both the HPT formula and of the observational
estimate).

I should remind at this point that the supergiant components in
some high mass X-ray binaries are significantly undermassive for
their luminosities (Zi\'o{\l}kowski 1977). In some systems, like
Cen X-3, this undermassivness is very serious and requires much
stronger mass loss than the normal stellar wind (Zi\'o{\l}kowski
1978). However, most likely, it is not the case for HDE 226868.
Its parameters may be fully explained by the evolution with the
normal stellar wind (as will be demonstrated in the further
discussion).

\subsection{The  evolutionary tracks}

Some of the obtained evolutionary tracks in the H-R diagram are
shown in Fig. 2. The careful reader might notice that the
luminosities of my models are by about 0.1 dex ($\sim$ 25\%)
smaller than the evolutionary tracks of Schalller  al. (1992).
This difference should be attributed mainly to the fact that I
used later edition of the opacity tables containing higher values
of the opacities (the opacity tables always evolve in the
direction of the growing opacity -- never the other way). The
correctness of the new opacities was confirmed by stellar
pulsations calculations -- see e.g. Pamyatnykh (1999). Part of the
luminosity difference is due to the fact, that I used higher
metallicity ($Z$ = 0.03 instead of $Z$= 0.02). And, finally, part
of the luminosity difference (all factors work in the same
direction) is due to the fact, that my stellar winds are stronger
(again due to higher metallicity).

As may be seen from a quick look at Fig. 2, the initial (zero age
main sequence) mass of HDE 226868 had to be in the range 35 to 55
M$_\odot$. The masses of the models corresponding to the present
day state of HDE 226868 (inside the $3 \sigma$ parallelogram) are
in the range 32.4 to 50.5 M$_\odot$. The central hydrogen content
in these models is between 0.126 and 0.265 and their evolutionary
age (since the beginning of the central hydrogen burning) is
between 2.7 and 4.3 $\times 10^6$ yr.

\begin{figure*}
 \begin{center}
%  \mbox{\epsfig{file=hrdcygx1.eps,angle=0,width=15.5cm}}
\centerline{\psfig{file=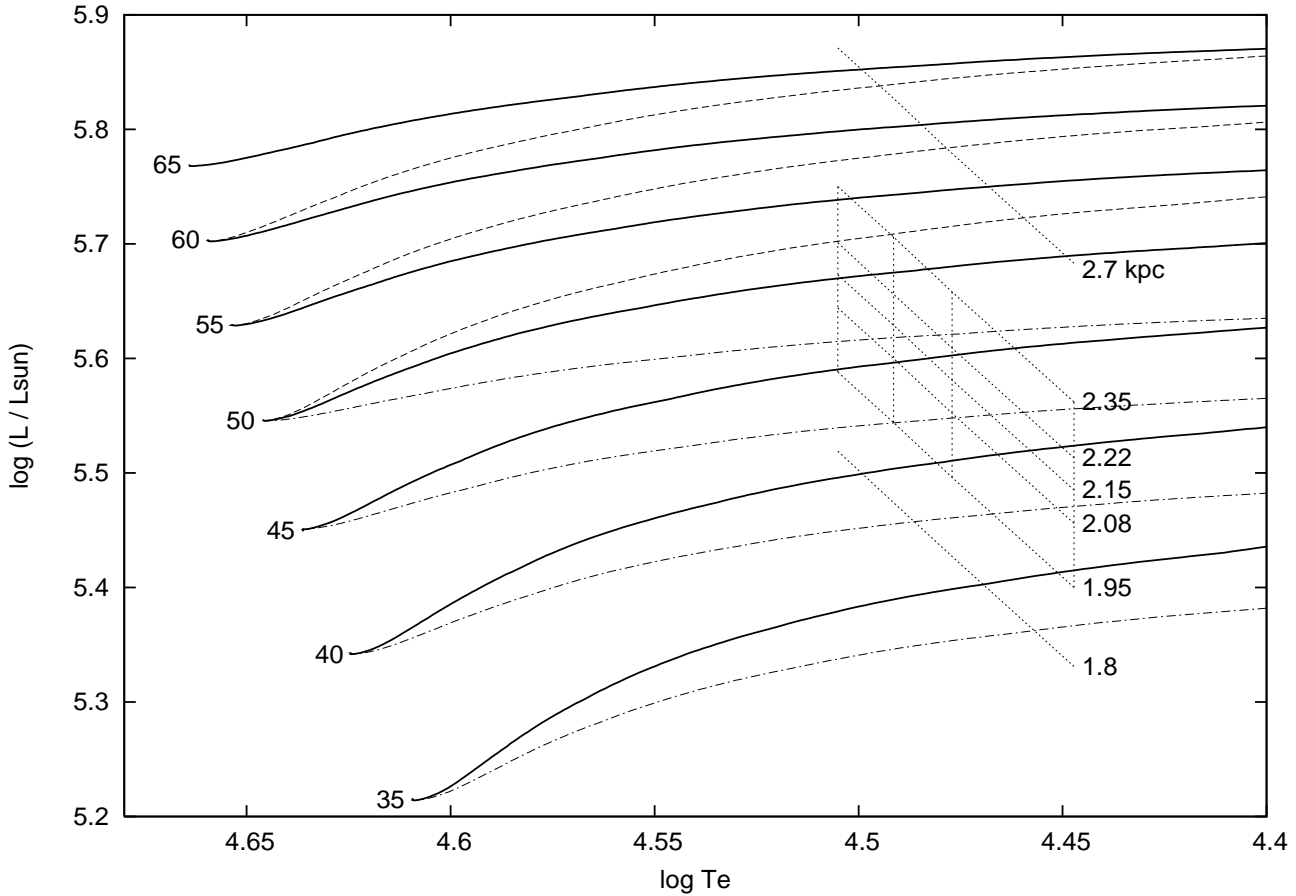,width=17.5cm}}
  \caption[h]{{\footnotesize The evolutionary tracks in the H-R diagram.
   The tracks are labeled with
   the initial mass of the star (in solar units). The solid lines
   describe the tracks computed with the stellar wind mass loss rates
   according to HPT (Hurley et al., 2000) formula. The broken lines
   and the dash-dotted lines describe the tracks computed with the
   mass loss rates smaller by a factor of two and larger by a factor
   of two, respectively. The slanted dotted lines correspond to the
   position of HDE 226868 for different assumed values of its distance
   (the assumed value of the distance in kpc is given
   at the right end of each line). The vertical dotted lines
   correspond to the effective temperatures of HDE 226868 equal
   (from left to right) to 32, 31, 30 and 28 $\times 10^3 $K. The most
   likely position of HDE 226868 lies within the large parallelogram
   ($\pm 3 \sigma$ error in distance).}}

 \end{center}
  \label{f2}
\end{figure*}

\begin{figure*}
 \begin{center}
%  \mbox{\epsfig{file=mld31.eps,angle=0,width=15.5cm}}
\centerline{\psfig{file=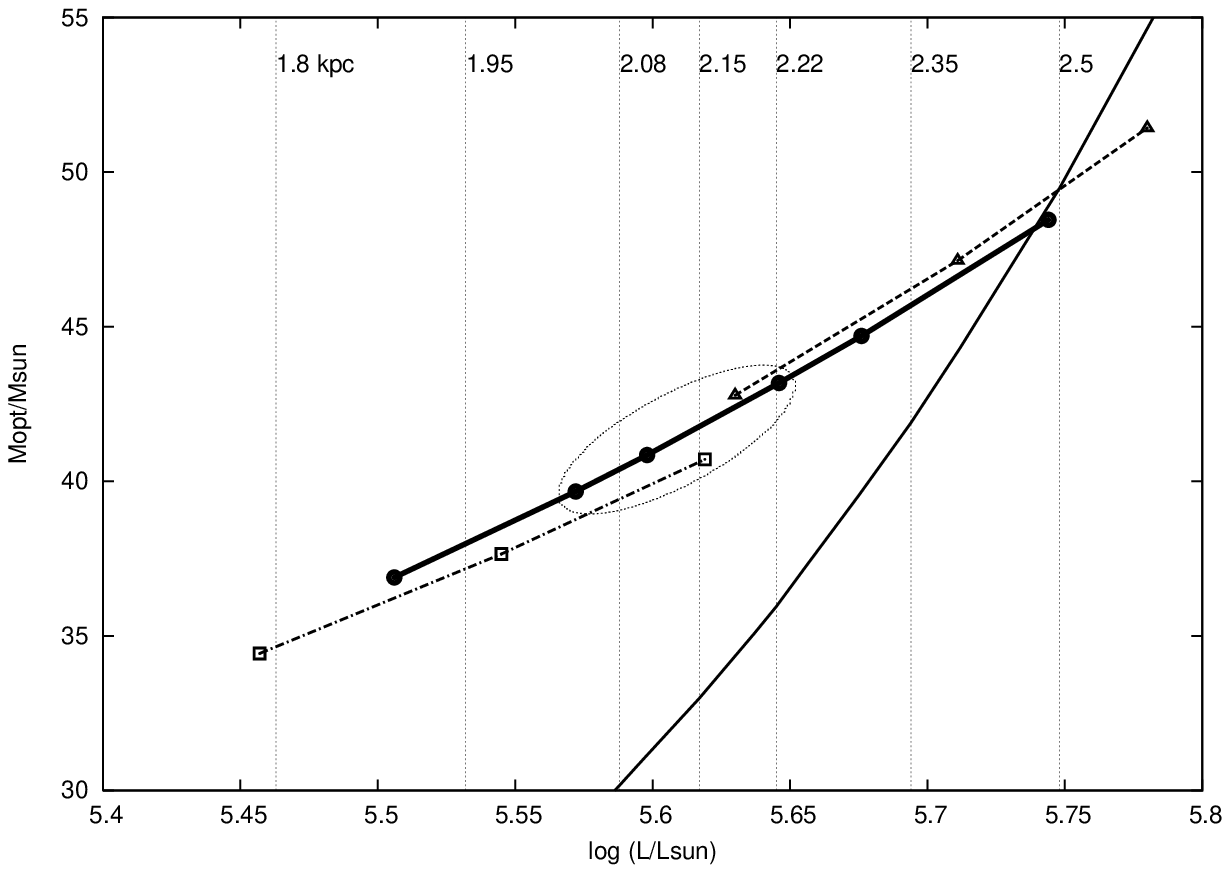,width=17.5cm}}
  \caption[h]{{\footnotesize The positions of the evolutionary models
   of HDE 226868 (with an assumed effective temperature $T_{\rm e}$ = 30700 K)
   in the mass$-$luminosity and mass$-$distance diagram ("mass" means here the
   present mass of the star). The scale of the distances (in kpc) is shown
   with the thin broken vertical lines (the most
   likely value of the distance is 2.15 $\pm$ 0.2 kpc, $3 \sigma$ error).
   The triangles, circles and squares correspond to the models
   with the stellar wind mass loss rates calculated with the HPT
   formula multiplied by the factors 0.5, 1 and 2, respectively.
   Only the models with the acceptable rates of mass loss (in the
   range 0.25 $\div$ 4 times the nominal observed value, $\dot{M} =
   - 2.6 \times 10^ {-6}$ M$_\odot$/yr) are shown. The solid line
   (without circles)
   describes the relation between the distance and the lower limit
   for the mass of HDE 226868 obtained in Section 2. The location of
   crossing of this relation by the sequences of the evolutionary
   models indicates that the distance to the binary system cannot be
   larger than $\sim$ 2.5 kpc (for the assumed effective
   temperature). The oval indicates the range of the models of the
   optical star for which viable models of the binary system could be
   constructed (see the discussion in the Section 5.2).}}
 \end{center}
  \label{f3}
\end{figure*}

To obtain evolutionary models which satisfactorily reproduce the
present day state of HDE 226868, one has to match not only the
luminosity and the effective temperature but also the rate of the
stellar wind mass loss. The first observational determination of
this parameter for HDE 226868 was done by Hutchings (1976) who,
analyzing the visual spectrographic data, got the value $\dot{M} =
- 2.5 \times 10^ {-6}$ M$_\odot$/yr. Persi et al. (1980) estimated
the rate of the mass outflow from the infrared emission of the
expanding circumstellar envelope and got the value $\dot{M}  = -
3.5 \times 10^ {-6}$ M$_\odot$/yr. More recent estimate of Herrero
et al. (1995) was based on the fits of the H$_\alpha$ profile and
led to the conclusion that the mass loss rate lies between 2 and 6
$\times10^ {-6}$ M$_\odot$/yr. They noted that the inaccuracy of
the fits is, probably, due to the fact that the stellar wind from
HDE 226868 is focused  towards the black hole (for the discussion
of the "focused" wind model see Gies \& Bolton 1986 and Miller et
al. 2002). In their test modelling of the atmosphere of HDE
226868, Herrero et al. were  using the value $\dot{M}  = - 4
\times 10^ {-6}$ M$_\odot$/yr and for their final model they chose
$\dot{M} = - 3 \times 10^ {-6}$ M$_\odot$/yr. The most recent
estimate by Gies et al. (2003) gave the value $\dot{M} = - 2.6
\times 10^ {-6}$ M$_\odot$/yr for the low/hard state (which is a
typical state of Cyg X-1) and $\dot{M} = - 2.0 \times 10^ {-6}$
M$_\odot$/yr for the high/soft state (which is less frequent in
this source). Taking all this into account, I assumed that the
observed rate of mass outflow from HDE 226868 is $\dot{M} = - 2.6
\times 10^ {-6}$ M$_\odot$/yr with an error by a factor of about
2, i.e. I assumed that the observed rate of mass outflow is
between 1.3 and 5.2 $\times 10^ {-6}$ M$_\odot$/yr. Let me remind
that our theoretical evolutionary models were using mass loss
rates in the range 0.5 to 2 times the rate given by HPT.
Altogether, it means, that the model for which the mass loss rate
calculated with the HPT formula would be by a factor of up to four
smaller or by a factor of up to four larger than the nominal
observational value ($\dot{M}  = - 2.6 \times 10^ {-6}$
M$_\odot$/yr) is still considered to be an acceptable match. This
is, probably, more than sufficient allowance for the uncertainty
of the mass loss rates. The parameters of some of these models are
given in Table 2. The positions of the acceptable models in the
mass$-$luminosity diagram (for the most likely value of the
effective temperature $T_{\rm e}$ = 30700 K) are shown in the Fig.
3. Since the luminosity of a model of HDE 226868 can be (for an
assumed effective temperature) directly translated into the
distance to the system (see Table 1), the Fig. 3 can be also
considered as the mass$-$distance diagram. The corresponding
distance scale is shown on this picture. The line corresponding to
the lower mass limit$-$distance relation obtained in section 2
(see Table 1 and Fig. 1) is also shown. One may notice that the
deduced mass of HDE 226868 depends mainly on the assumed distance
to the binary system, and only very weakly on the assumptions
about the mass loss rates. For the assumed effective temperature
($T_{\rm e}$ = 30700 K) this mass has to be in the range 35 $\div$
45 M$_\odot$ (if the distance to the system is in the range 1.8 to
2.35 kpc). One may also conclude (with the help of the lower mass
limits derived in Section 2) that the distance to the system
cannot be larger than $\sim$ 2.5 kpc (still for the assumed
effective temperature).

Similar diagrams as Fig. 3 may be constructed for other possible
values of the effective temperature of HDE 226868. Qualitatively,
they look similar to the Fig. 3. For the effective temperature
$T_{\rm e}$ = 28000 K the mass of the star has to be in the range
29 $\div$ 35 M$_\odot$ and the upper limit for the distance to the
system is $\sim$ 2.1 kpc. For the effective temperature $T_{\rm
e}$ = 32000 K the mass of the star has to be in the range 37
$\div$ 50 M$_\odot$ and the upper limit for the distance to the
system is $\sim$ 2.7 kpc.

\section{Discussion}

\subsection{The  mass of HDE 226868}

The obvious and fairly strong conclusion derived from the
evolutionary calculations is that HDE 2226868 had to be quite
massive in the past and is still very massive at present. To put
it very briefly, it has to be very massive because it is very
bright (while still burning hydrogen in the core). For the most
likely range of parameters (30000 to 31000 K for the effective
temperature and 2.08 to 2.22 kpc for the distance) the present
mass of HDE 2268668 has to be in the range 37 to 44 M$_\odot$.
Increasing the interval of the possible values of the distance to
1.95 to 2.35 kpc increases this range only slightly (to 33.5 $\div
47 M_\odot$). Extending the interval of the possible values of the
effective temperatures to 28000 to 32000 K leads to the range of
possible masses 31 $\div 50 M_\odot$. Finally, extending the range
of the possible distances down to 1.8 kpc (the smallest value
quoted in the literature) results in the final range of the
possible masses of HDE 226868: 29 $\div 50 M_\odot$. The value of
29 M$_\odot$ seems to be a firm lower limit for the present mass
of HDE 226868.

One may ask whether it is possible to construct substantially less
massive evolutionary models by taking more liberal estimate of the
observational uncertainties or accounting for some simplifications
of our models. The answer is negative. Decreasing the effective
temperature of HDE 226868 would result in lower luminosities and
so would lead to less massive configurations. However, the value
$T_{\rm e}$ = 28000 K, taken as the lower limit to the effective
temperature in the above consideration is probably already too low
(the true value is probably in the range 30000 to 31000 K).
Decreasing the minimum distance from 1.95 to 1.8 kpc decreased the
minimum possible masses of the models by only $\sim 2 M_\odot$, as
was mentioned above. Let us now consider the possible consequences
of some simplifications of our evolutionary calculations. I
neglected the possible dumping of matter on HDE 226868 from the
progenitor of Cyg X-1 (and so "rejuvenation" of the star) during
the first mass loss/exchange phase. Most likely, there was no
significant accretion on HDE 226868 because any substantial mass
transfer in such a close binary and in so early stage of evolution
(when both stars had no distinct cores) would lead to the rapid
built-up of a common envelope and the coalescence of both
components. So, most likely, the first mass transfer was mainly
the mass loss from the system. If there were no serious accretion,
then the nuclear evolution prior to the mass transfer is properly
accounted for in our calculations. If (which is rather unlikely)
there was a substantial mass dumping on HDE 226868, then the
nuclear evolution before the mass transfer is negligible because
the substantially less massive star was evolving much more slowly
(large mass gain essentially resets the evolutionary clock to
ZAMS).

One may wonder whether smaller metallicity ($Z$ = 0.02 instead of
$Z$= 0.03) would not be a better choice for HDE  226868. To check
the possible effects of such alternative choice, I calculated
several evolutionary tracks  for stars with $Z$ = 0.02. As might
be expected, the changes in the results were small. For the models
reproducing the "best" parameters of HDE 226868, the masses
decreased from about 41.7 M$_\odot$ to about 39.9 M$_\odot$ (only
about 4\%).

There remains a rather remote possibility that the chemical
composition of HDE 226868 is not normal, i.e. that it contains
less hydrogen than a normal Population I star. E.g. Herrero et al.
(1995) concluded, from fitting models of the atmosphere, that
helium might be overabundant by a factor of about two. The
evolutionary calculations presented here indicate that the stellar
wind mass loss from HDE 226868 was not severe enough for hydrogen
depleted layers to show up on the surface. However, if a large
amount of matter with significantly decreased hydrogen content was
dumped on HDE 226868 from the progenitor of Cyg X-1, then the star
might contain less hydrogen than the assumed value of 0.7. The
lack of the CNO anomalies (Dearborn 1977) testifies against this
being the case. However, to check the consequences of such (rather
unlikely) situation, I calculated several evolutionary sequences
for the stars with the initial chemical composition $X$=0.5,
$Z$=0.03 (this corresponds to an overabundance of helium by a
factor of two). As could be easily expected, the models with
similar luminosities were now less massive, but the change was not
very substantial. For the "best fit" models, the masses decreased
from about 41.7 M$_\odot$ to about 35.3 M$_\odot$ (only about
15\%). However, I would like to stress again, that there are no
good reasons to expect such abnormal hydrogen content in HDE
226868.

The only way to produce an evolutionary model with the mass $\la
20$ M$_\odot$ is to decrease drastically the distance to the
system. I constructed one evolutionary track that produced a 19.5
M$_\odot$ configuration at the effective temperature 30700 K with
the mass loss rate $\dot{M}  \approx - 2.5 \times 10^ {-6}$ (the
initial mass of the star was 32.5 M$_\odot$). However, the price
was very high: the distance had to be decreased to $\sim$ 0.9 kpc
and the rate of the mass loss during the evolution had to be
multiplied by a factor of $\sim$ 20 with respect to the HPT
formula. Any of these conditions, even taken separately, seems
extremely unlikely.

There is a clear conflict between the above considerations and the
mass estimate given by Herrero et al. (1995). They used atmosphere
models to reproduce the observed spectrum of HDE 226868. Then,
they used their fit to determine the effective temperature, the
surface gravity, the radius and the helium abundance of the star.
Their mass estimate ($M \sim 18 \pm 4$ M$_\odot$) is a consequence
of the surface gravity and the radius determination.

I see no solution of this conflict. I may only repeat that the
evolutionary calculations for such an early evolutionary phase are
very robust and rather difficult to dispute (it is, essentially,
the mass-luminosity relation for the main sequence stars). I may
also add that the evolutionary results presented above are roughly
consistent with the model of Gies \& Bolton (1986a), based on an
extensive analysis of the large and diversified collection of the
observational data.

\subsection{The  Mass of Cyg X-1}

In the previous section we concluded that the present mass of HDE
226868 has to be in the range of 29 $\div $ 50 M$_\odot$ (assuming
the widest possible ranges of the distances and the effective
tempeatures). The mass of its companion, black hole Cyg X-1, does
not come out directly from the evolutionary calculations. However,
it can be calculated (with the help of some observational
constraints), once we select a chosen evolutionary model of HDE
226868. Once the selection is made, we know the mass $M_{\rm opt}$
and the radius $R_{\rm opt}$ of the optical component. We can also
calculate the distance, with the help of eq. (1) or (2) (or the
corresponding expressions for other effective temperatures).
Subsequently, we can use two equations to solve for the
inclination of the orbit $i$ and the mass ratio $q = M_{\rm
opt}/M_{\rm x}$. One of these equations makes use of the mass
function,

\begin{eqnarray}
f(M_{\rm x}) = M_{\rm opt} {\rm sin}^3 i/[q(1+q)^2].
\end{eqnarray}

\noindent The other relates the radius of the star to the size of
the orbit,

\begin{eqnarray}
R_{\rm opt}& =& R_{\rm RL} \times f_{\rm RL} =\nonumber\\ &=&
f_{\rm RL} (0.38 + 0.2 \log\hspace*{.5ex} q) A =\nonumber \\ &=&
f_{\rm RL} (0.38 + 0.2 \log\hspace*{.5ex} q) a_1 (1+q),
\end{eqnarray}

\noindent where $R_{\rm RL}$ is the radius of the Roche lobe
around HDE 226868, $f_{\rm RL}$ is the fill-out factor ($f_{\rm
RL} = R_{\rm opt}/R_{\rm RL}$), $A$ is the orbital separation of
the binary components and $a_1$ is the radius of the orbit of HDE
226868 (see section 2).

Inserting the observational data, eqs. (3)$-$(4) can be written
as:

\begin{eqnarray}
M_{\rm opt} {\rm sin}^3 i/[q(1+q)^2] = 0.251,
\end{eqnarray}

\begin{eqnarray}
R_{\rm opt} = f_{\rm RL} (0.38 + 0.2 \log \hspace*{.5ex} q) (1+q)
\times 8.36/{\rm sin}\hspace*{.5ex} i.
\end{eqnarray}

Once a given evolutionary model of HDE 226868 is selected from the
grid of the acceptable models and a value of the parameter $f_{\rm
RL}$ is assumed, the eqs. (5)$-$(6) can be solved for $i$ and $q$.
Knowing $q$ we can immediately calculate also the mass of compact
component $M_{\rm x}$. In principle, this procedure can be applied
to any combination of the evolutionary model and of the value of
$f_{\rm RL}$. In fact, however, not every evolutionary model of
HDE 226868 (acceptable if we consider the optical component alone)
permits the construction of a consistent model of the binary
system. This is because of the observational constraints on the
value of the inclination $i$ and, especially, because of the
strong observational constraints on the value of the fill-out
factor $f_{\rm RL}$. As demonstrated by Gies \& Bolton (1986a,b),
in order to explain quantitatively the He emission lines produced
in the stellar wind from HDE 226868, the fill-out factor $f_{\rm
RL}$ has to be larger than 0.9 and, most likely, not smaller than
0.95 (perhaps the best value would be around 0.98). On the other
hand, Gies \& Bolton demonstrated that observed rotational
broadening of the photospheric absorption lines of HDE 226868 and
the observed amplitude of the ellipsoidal light variations impose
substantial constraints on the values of the mass ratio, the
fill-out factor and the inclination. The rotational broadening
depends mainly on the mass ratio and the measured broadening
indicates the mass ratio in the range $\sim 2 \div 2.5$. The
amplitude of the ellipsoidal light variations is determined mainly
by the fill-out factor and the inclination. For the assumed values
of $f_{\rm RL}$ equal 0.9, 0.95 and 1, the resulting inclination
is $\sim 38^{\rm o}$, $\sim 33^{\rm o}$ and $\sim 28^{\rm o}$,
respectively.

Constructing models of the binary system to be consistent with
observations, I assumed that for any adopted value of $f_{\rm
RL}$, the calculated inclination should be within $\pm 5^{\rm o}$
from the corresponding values quoted above. I started with
different models of the optical component, acceptable from the
point of view of the stellar evolution, as described in section
4.3. Then, I assumed the value of $f_{\rm RL}$ equal 0.95 and
solved eqs. (5)$-$(6) to find $q$, $i$ and $M_{\rm x}$.
Subsequently, I tried higher values of $f_{\rm RL}$. The higher
values of $f_{\rm RL}$ produced the solutions with lower (in many
cases unacceptably low) values of the inclination. The dependence
of $q$, $i$ and $M_{\rm x}$ on the assumed value of $f_{\rm RL}$
may be seen from many sequences of binary models, presented in
Tab. 2. For all sequences included in the table, I present only
two limiting models for the lowest and the highest value of
$f_{\rm RL}$ for which an acceptable model could still be
obtained. If there is only one entry for a given sequence (as is
the case for $M_{\rm opt} = 39.67$ M$_\odot$  and $M_{\rm opt} =
47.37$ M$_\odot$), it means that only one value of $f_{\rm RL}$
produced an acceptable solution (the value of $f_{\rm RL}$ was
varied with the increment of 0.01).

I classified, as acceptable, the models satisfying the following
criteria:

\hspace{1cm} (1) $d = 1.8 \div 2.35$ kpc

\hspace{1cm} (2) $T_{\rm e} = 28000 \div 32000$ K

\hspace{1cm} (3) $-\dot{M}  = 1.3 \div 5.2 \times 10^ {-6}$
M$_\odot$/yr

\hspace{1cm} (4) $f_{\rm RL} \ge 0.95$

\hspace{1cm} (5) $i = 28^{\rm o} + (1-f_{\rm RL})\times 100^{\rm
o} \pm 5^{\rm o}$ (this corresponds to $i \approx 28 \div 38^{\rm
o}$ for $f_{\rm RL} = 0.95$ and $ i \approx 23 \div 33^{\rm o}$
for $f_{\rm RL} = 1.00$, as advocated by Gies \& Bolton 1986a).

Parameters of selected acceptable models are given in the second
part of Table 2. For each assumed value of the effective
temperature, I present two sequences of models corresponding to
the lowest and the highest current mass of the optical component.
The exception to this rule is the sequence of models for $T_{\rm
e} = 30700$ K and $M_{\rm opt} = 41.67$ M$_\odot$ ($d$ = 2.15
kpc). The initial evolutionary parameters ($M_0$ and $f_{\rm SW}$)
of the optical component model for this sequence were adjusted so
as to obtain the perfect fit with the "best" values of the
observational parameters of HDE 226868 (effective temperature,
luminosity  and the rate of stellar wind mass loss). All models
from this sequence (for $f_{\rm RL} = 0.96 \div 1.00$) are
successful from the point of view of our criteria.

There are several conclusions that can be drawn from the
collection of the obtained models of the binary system (only some
of these models are shown in Table 2). The first concerns the mass
of the compact component. Assuming the widest possible intervals
of the distances and the effective temperatures: 1.8 to 2.35 kpc
and 28000 to 32000 K, the mass of the compact component must be in
the range 13.5 $\div$ 28.5 M$_\odot$. For the most likely
intervals of these parameters: 1.95 to 2.35 kpc and 30000 to 31000
K, this range narrows to 15.5 $\div$ 25 M$_\odot$. The second
conclusion is related to the distance to the binary system. It
appears that for a given (assumed) effective temperature,
consistent models are possible only for some, relatively narrow,
interval of distances (the distances out of this interval would
require unacceptably low or unacceptably high values of the
inclination). For the effective temperatures equal to 28, 30,
30.7, 31 and 32 $\times 10^3 $K, the corresponding intervals are
1.80$\div$1.88, 1.98$\div$2.15, 2.04$\div$2.22, 2.07$\div$2.27 and
2.21$\div$2.35 kpc, respectively. It is certainly encouraging that
the most likely range of the effective temperatures (30000 to
31000 K) requires the distance interval (1.98$\div$2.27 kpc) that
almost exactly coincides with the independent estimate of the most
likely distance range (1.95$\div$2.35 kpc). The obtained relation
between the distance and the effective temperature of HDE 226868
means that, if in the future the distance will be known more
precisely (e.g. from the future astrometric space missions), then
it will be possible to set constraints on the effective
temperature. For example, if it is found that the distance to the
system is greater than 2.1 kpc, it would mean that the effective
temperature of HDE 226868 has to be higher than 30000 K. Also the
opposite is true. If it is found (e.g. from the better models of
the atmospheres) that the effective temperature of HDE 22868 is
greater than 30000 K, it would mean that the distance to the
system must be larger than 2.0 kpc (and this conclusion would not
depend on any photometric or astrometric considerations).

Finally, the third conclusion confirms the earlier results (based
on the evolutionary calculations alone) limiting the present mass
of HDE 226868 to the range $29 \div 50$ M$_\odot$ and the initial
mass to the range $33.5 \div 55$ M$_\odot$.

Let us note that for the black hole mass in the range 15 $\div$ 25
M$_\odot$, the state transitions occurs in Cyg X-1 at the
luminosity level equal 0.025 to 0.015 of the Eddington luminosity
(assuming $d$ = 2.15 kpc and using the flux values given by
Zdziarski et al. 2002). These values lie in the range 0.007 to
0.03 found for other X-ray binaries (Maccarone 2003).

To summarize this section, the evolutionary considerations provide
much stronger lower limits to the masses of both components than
the model independent analysis, presented in section 2. In
addition, they yield some independent constraints on the distance
to the system.

\section{The Possible Evolutionary Scenario}

Let us speculate a little bit about the possible evolutionary past
of our binary system. It had to start as a very massive system
(the initial primary had to complete its evolution by the time the
secondary (HDE 226868) reached its present evolutionary state i.e.
in less than $\sim 3 \div 4 \times 10^6$ yr). The initial masses
of the components were probably $\sim 80 \div 100$ and $\sim 40
\div 50$ M$_\odot$. The massive primary (the progenitor of a black
hole) was shedding the mass mainly in the form of the stellar wind
(there was probably little, if any, mass transfer to the
companion). The mass of the primary decreased, at the end of its
evolution (before the collapse), to $\sim 15 \div 25$ M$_\odot$.
The low spatial velocity of Cyg X-1 with respect to its parental
association indicates, as argued by Mirabel \& Rodriguez (2003),
that the final collapse to a black hole proceeded with very little
($\la 1$ M$_\odot$) mass ejection. It could be, even, a prompt
collapse with no accompanying supernova explosion at all
("formation of a black hole in the dark"). The secondary (the
progenitor of HDE 226868) was also losing mass in the form of the
stellar wind, although at a lower rate. This mass loss took place
both before and after the collapse of its companion. This process
decreased the mass of the secondary to the present value of $\sim
35 \div 45$ M$_\odot$. In this way, the present day binary
consisting of a massive supergiant of $\sim 40 \pm 5$ M$_\odot$
(with a central hydrogen content of $\sim 0.2 \div 0.3$) and a
black hole of $\sim 20 \pm 5$ M$_\odot$ was formed.

The above evolutionary history is only only one of the possible
scenarios. As noted by Gies (2004, private communication), the
above scenario implies that about half of the total initial mass
of the binary system was lost in the form of stellar wind. This
implies, in turn, that the initial orbital period was $\la 3$
days, which means that the system was very tight. One cannot
exclude that the initial masses were smaller, the initial orbital
period longer and the system passed through the mass transfer and
the common envelope phase.

\section{Conclusions}

\tem{(1)} The calculations modelling the evolution of HDE 226868,
under different assumptions about the stellar wind mass loss rate,
provide robust limits on the present mass of the star. For the
most likely intervals of the values of the distance and of the
effective temperature: 1.95 to 2.35 kpc and 30000 to 31000 K, the
mass of HDE22868 is 40 $\pm5$ M$_\odot$. Extending the intervals
of these parameters to 1.8 to 2.35 kpc and 28000 to 32000 K, one
obtains the mass of the star in the range $29 \div 50$ M$_\odot$.

\tem{(2)} Including the additional constraints resulting from the
observed properties of the binary system HDE 226868/Cyg X-1, one
can estimate the mass of the black hole component. For the most
likely values of the parameters mentioned in the item (1) above
this mass is 20 $\pm5$ M$_\odot$. For the extended intervals of
the parameters the mass is in the range of $\sim 13.5 \div 29$
M$_\odot$.

\tem{(3)} The distance to the binary system has to be in the
ranges 1.80$\div$1.88, 1.98$\div$2.15, 2.04$\div$2.22,
2.07$\div$2.27 and 2.21$\div$2.35 kpc, for the effective
temperature of HDE 226868 equal to 28, 30, 30.7, 31 and 32 $\times
10^3$ K, correspondingly.

\section*{Acknowledgements}

I would like to thank A. Zdziarski for careful reading of the
manuscript and for many helpful comments and stimulating
discussions. I would like also to thank D. Gies who, acting as a
referee, made several comments and suggestions which helped to
improve this paper. This work was partially supported by the State
Committee for Scientific Research grants No 1 P03D 018 27 and No
PBZ KBN 054/P03/2001.

\begin{table*}
%\centerline{\bf Tab. 3 $-$ Parameters of the selected evolutionary
%models} \nobreak \centerline{\bf of the binary system HDE
%226868/Cyg X-1} \nobreak \vspace{5mm}
%\nopagebreak
\centering
 \begin{minipage}{140mm}
  \caption{Parameters of the selected evolutionary models of the
   binary system HDE 226868/Cyg X-1.}

%\moveleft 12mm
\vbox{
\begin{tabular}{|l|r|r|l|l|r|r|r|r|r|r|}
\hline
%&&&&&&&&&&\\
\multicolumn{1}{|c|}{$T_{\rm e}$}&\multicolumn{1}{|c|}{$M_{\rm opt}$}&\multicolumn{1}{|c|}{$d$}&\multicolumn{1}{|c|}{$M_0$}&\multicolumn{1}{|c|}{$f_{\rm SW}$}&\multicolumn{1}{|c|}{$R_{\rm opt}$}&\multicolumn{1}{|c|}{log$L$}&\multicolumn{1}{|c|}{$-\dot{M}$}&\multicolumn{1}{|c|}{$f_{\rm RL}$}&\multicolumn{1}{|c|}{$i$}&\multicolumn{1}{|c|}{$M_{\rm x}$}\\
\multicolumn{1}{|c|}{[10$^3$
K]}&\multicolumn{1}{|c|}{[M$_{\odot}$]}&\multicolumn{1}{|c|}{[kpc]}&\multicolumn{1}{|c|}{[M$_{\odot}$]}&&\multicolumn{1}{|c|}{[R$_{\odot}$]}&
\multicolumn{1}{|c|}{[L$_{\odot}$]}&\multicolumn{1}{|l|}{[10$^{-6}$
}&&\multicolumn{1}{|c|}{[$^{\rm
o}$]}&\multicolumn{1}{|c|}{[M$_{\odot}$]}\\
&&&&&&&\multicolumn{1}{|r|}{M$_{\odot}$/yr]}&&&\\
%&&&&&&&&&&\\
\hline \multicolumn{11}{|c|}{}\\ \multicolumn{11}{|c|}{Observational parameters}\\
\multicolumn{11}{|c|}{}\\ \hline
%&&&&&&&&&&\\
30.7&&2.15&&&22.77&5.617&2.60&$\ge$ 0.95&33&\\
-2.7,+1.3&&$\pm$0.2&&&$\pm$2.3&$\pm$0.13&-1.3,+2.6&&$\pm$5&\\
%&&&&&&&&&&\\
\hline \multicolumn{11}{|c|}{}\\
\multicolumn{11}{|c|}{Acceptable models}\\
\multicolumn{11}{|c|}{}\\ \hline
%&&&&&&&&&&\\
28&29.30&1.80&33.5&2&19.69&5.330&1.84&0.95&34.6&13.6\\
&&&&&&&&0.98&25.2&19.9\\
28&30.31&1.88&35&2&20.53&5.367&2.12&0.98&33.9&14.2\\
&&&&&&&&1.00&27.7&18.0\\
30&36.74&1.98&40&1&21.10&5.511&1.68&0.95&30.0&18.3\\
&&&&&&&&0.96&27.0&20.7\\
30&39.53&2.15&44.3&1.17&22.88&5.581&2.60&0.99&33.9&16.6\\
&&&&&&&&1.00&30.7&18.5\\
30.7&39.67&2.04&43.5&1&21.63&5.572&2.07&0.95&28.9&19.9\\
&&&&&&&&&&\\ {\bf 30.7}&{\bf 41.67}&{\bf 2.15}&{\bf 46.3}&{\bf
1.05}&{\bf 22.77}&{\bf 5.617}&{\bf 2.60}&{\bf 0.96}&{\bf
36.1}&{\bf 16.0}\\ &&&&&&&&{\bf 1.00}&{\bf 24.1}&{\bf 25.6}\\
&&&&&&&&&&\\
30.7&43.18&2.22&48&1&23.56&5.647&2.78&0.99&32.8&18.1\\
&&&&&&&&1.00&29.7&20.3\\
31&40.84&2.12&50&2&22.37&5.618&4.88&0.95&36.1&15.8\\
&&&&&&&&0.99&24.0&25.4\\
31&44.80&2.27&50&1&23.88&5.675&3.07&0.99&32.7&18.6\\
&&&&&&&&1.00&29.6&20.8\\
32&47.37&2.21&50&0.5&23.13&5.702&1.63&0.95&29.2&21.8\\
&&&&&&&&1.00&23.6&28.6\\
32&50.50&2.35&53.6&0.5&24.54&5.754&2.01&0.96&37.3&17.3\\
&&&&&&&&1.00&25.1&27.0\\
%&&&&&&&&&&\\
\hline
\end{tabular}}

\vspace{4mm}
{\footnotesize NOTES:\vspace{2mm}\\
%\vspace{4mm}
(1) $M_0$ denotes the initial (ZAMS) mass of the optical
component, $f_{\rm SW}$ denotes the multiplying factor applied to
HPT formula; other symbols have their usual meanings\\
(2) The bold face entries correspond to the "best fit" models (see
the text)\\}
\vspace{5mm}
\end{minipage}
\end{table*}
%\vspace{10mm} \nopagebreak

%\vspace{15mm}

\label{lastpage}
\end{document}